\documentclass[aps,pra,showpacs]{revtex4}

\usepackage{amssymb}
\usepackage{amsmath}
\usepackage[xdvi,dvips]{graphicx}
\usepackage{fancybox, fancyhdr, amsthm, verbatim}
\usepackage[all]{xy}

\usepackage{graphicx}% Include figure files
\usepackage{subfigure}

\newcommand{\br}{{\bf{r}}}
\newcommand{\bk}{{\bf{k}}}

\newcommand{\prt}{\partial}

\newcommand{\om}{\omega}
\newcommand{\eps}{\varepsilon}

\newcommand{\la}{\lambda}

\begin{document}

\title{Stationary wave patterns generated by an impurity moving
with  supersonic velocity through a Bose-Einstein condensate}

\author{T.-L. Horng$^{1}$}
\author{S.-C. Gou$^{2}$}
\author{T.-C. Lin$^3$}
\author{G.A. El$^4$}
\author{A.P. Itin$^{5,6}$}
\author{A.M. Kamchatnov$^7$}

\affiliation{
$^{1}$ Department of Applied Mathematics, Feng Chia University, Taichung 40724, Taiwan\\
$^{2}$ Department of Physics, National Changhua University of Education, Changhua 50058, Taiwan \\
$^{3}$ Department of Mathematics, National Taiwan University, Taipei, Taiwan \\
$^{4}$ Department of Mathematical Sciences, Loughborough
University, Loughborough LE11 3TU, UK \\
$^{5}$ Applied Physics Department, Helsinki University of Technology,
P.O. Box 5100, 02015, Finland \\
$^{6}$ Space Research Institute, Russian Academy of Sciences, 
       Moscow, Russia    \\
$^7$ Institute of Spectroscopy, Russian Academy of Sciences,
Troitsk, Moscow Region, 142190, Russia
}

\date{\today}

\begin{abstract}
Formation of stationary 3D wave patterns generated by a small point-like impurity moving through a Bose-Einstein
condensate with supersonic velocity is studied. Asymptotic formulae for a stationary far-field density distribution are obtained. Comparison with three-dimensional
numerical simulations demonstrates that these formulae are accurate enough already
at distances from the obstacle equal to a few wavelengths.
\end{abstract}


\maketitle

\section{Introduction}

As is well known, superfluidity means that slow enough flow of a fluid is not accompanied by heat
production or generation of excitations of any kind. As a result, movement of a fluid is free
of dissipation. In a similar way, motion of impurity through a superfluid goes on without any
friction for small enough values of its velocity. The threshold velocity above
which superfluidity is lost
is determined by various physical mechanisms depending on the nature of the fluid and geometry
of the process. For example, in the original Landau theory \cite{landau1,landau2} of superfluidity
it breaks down when the generation of rotons becomes possible which leads to a famous Landau criterion
for superfluidity. However, Landau's estimate for this mechanism of dissipation gives too large
threshold velocity for HeII, and this disagreement with the theory was explained by Feynman
\cite{feynman} by the possibility of the generation of vortex rings. This phenomenon is essential for
large enough obstacles with the size about the healing length. In Bose-Einstein condensates
(BECs) of rarefied
gases the healing length can be relatively large and generation of vortices by small impurities
becomes ineffective. In this case BEC remains superfluid for all
velocities less than the minimal sound velocity corresponding to the long wavelength limit
of the Bogoliubov dispersion law. For a supersonic motion of an impurity, the Cherenkov
radiation of sound waves is the main mechanism of the appearance of friction and the corresponding
``drag force'' was calculated in \cite{km-01,ap-04}.

However, the detailed wave pattern generated by a moving impurity is also of a considerable
interest. This problem became very topical in two-dimensional (2D) case in connection with
the results of the experiment \cite{cornell05,carusotto} in which the waves were generated by
the flow of a condensate expanding through an obstacle created by a laser beam. Since such
an obstacle has the size comparable or greater than the healing length, the arising
here wave pattern can be quite complicated. Already in numerical experiment \cite{wmca-99}
modeling a similar situation
it was noticed that the interference of sound (Bogoliubov) waves yields the wave pattern
located outside the Mach cone. Analytic theory of such wave patterns was developed in
\cite{carusotto,gegk,gsk}. In many respects, this theory is analogous to the well-known
Kelvin's theory of ``ship waves'' generated by a ship moving in still deep water,
with the dispersion law for the surface water waves replaced by the Bogoliubov dispersion
law for sound waves in BEC. Besides that, due to a large size of the laser beam,
vortices or oblique dark solitons located inside the Mach cone can also be generated
by a 2D flow of a BEC. The corresponding theory was developed in \cite{egk1,egk2,kp08} and
was recently generalized  \cite{two-comp} for a two-component condensate.  However
analogous theory for 3D flow has not been developed yet, although it is of
considerable interest for understanding of the wave processes in BECs (see, e.g., \cite{sh-08}). 
In this paper, we shall consider both analytically and numerically the 3D wave pattern 
created by a small impurity moving with supersonic velocity through a bulk Bose-Einstein
condensate.

\section{Stationary wave pattern}

Dynamics of BEC of rarefied gases is described very well by the Gross-Pitaevskii (GP)
equation
\begin{equation}\label{GP}
    i\psi_t+\tfrac12\Delta\psi +(1-|\psi|^2)\psi-V\psi=0 \, ,
\end{equation}
which is written here in standard non-dimensional notation (see, e.g., \cite{gsk}),
so that the density of an undisturbed BEC with the repulsive interaction between atoms
is equal to unity. We suppose that the external potential $V$ is created by a
point-like impurity moving with velocity $\mathbf{U}$ along $x$ axis in negative direction,
\begin{equation}\label{2-2}
    V(\br)=V_0\delta(\br+\mathbf{U}t).
\end{equation}
The stationary wave pattern can be obtained for a supersonic velocity $\mathbf{U}$
which in our non-dimensional units with the sound velocity equal to unity means
\begin{equation}\label{2-3}
    |\mathbf{U}|\equiv M>1,
\end{equation}
$M$ being the Mach number. Assuming that the interaction with impurity is small,
we can apply the perturbation theory \cite{km-01,ap-04,gsk} and linearize
Eq.~(\ref{GP}) with respect to small disturbance $\delta\Psi$ of the wave function,
$\Psi=1+\delta\Psi$. Then $\delta\Psi$ satisfies the equation
\begin{equation}\label{2-4}
    i\delta\Psi_t+\tfrac12\Delta\delta\Psi-(\delta\Psi+\delta\Psi^*)-
    V_0\delta(\mathbf{r}+\mathbf{U}t)=0
\end{equation}
which can be readily solved by the Fourier method and the resulting perturbation
of density $\delta n=\delta|\Psi|^2\cong\delta\Psi+\delta\Psi^*$ is given by the
expression (see, e.g., Eq.~(20) in \cite{gsk})
\begin{equation}\label{3-1}
    \delta n=V_0\int\frac{k^2e^{i\mathbf{k}\mathbf{r}}}
    {(\mathbf{k}\mathbf{U})^2-k^2(1+k^2/4)+i\eps}
    \frac{d^3k}{(2\pi)^3},
\end{equation}
where $\eps=|\eps|\cdot\mathrm{sgn}(k)$ is an infinitely small parameter, $|\eps|\to0$,
which determines the rule of going around the poles of the integrand function.

\begin{figure}[bt]
\begin{center}
\includegraphics[width=6cm]{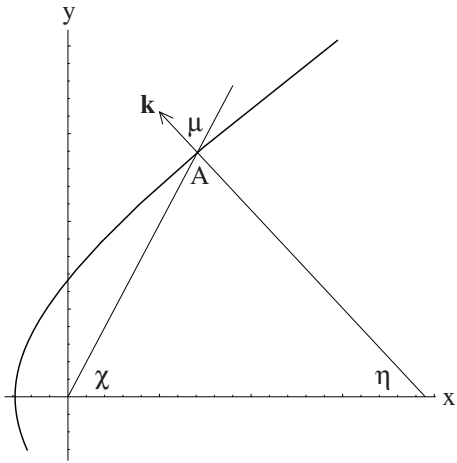}
\caption{Coordinates defining the radius-vector $\mathbf{r}$ and the
wave vector $\mathbf{k}$. The latter one is normal to the wave crest line
of the ship-wave which is shown schematically by a curve.}
\end{center}\label{fig6}
\end{figure}

Since the wave pattern is axially symmetric with respect to the $x$ axis, it is convenient
to define the coordinate system so that the observation point lies in the $(x,y)$ plane.
Then vector $\br$ has the components
\begin{equation}\label{3-2}
    \br=(r\cos\chi,r\sin\chi,0),
\end{equation}
where $\chi$ is the polar angle between $\br$ and $x$ axis. Let the vector $\bk$ lie
in the plane making an angle $\phi$ with the $(x,y)$ plane. Then its components
can be parameterized as
\begin{equation}\label{3-3}
    \bk=(-k\cos\eta,k\sin\eta\cos\phi,k\sin\eta\sin\phi),
\end{equation}
where $k\cos\eta$ is the projection of vector $\bk$ on the $x$ axis.
The geometrical meaning of the angles $\eta$, $\chi$, and $\mu=\pi-\eta-\chi$
is shown in Fig.~1.

Substitution of Eqs.~(\ref{3-2}) and (\ref{3-3}) into Eq.~(\ref{3-1}) and
simple transformations cast this
expression into the form
\begin{equation}\label{3-4}
    \delta n=\frac{V_0}{\pi^2}\int_0^\infty\int_0^{\pi}
    \frac{k^2\sin\eta J_0(kr\sin\eta\sin\chi)e^{-ikr\cos\eta\cos\chi}}
    {k^2-k_0^2-i\eps}dk d\eta \, ,
\end{equation}
where
\begin{equation}\label{3-5}
    k_0=2\sqrt{M^2\cos^2\eta-1}
\end{equation}
and we have used the well-known integral representation
\begin{equation}\label{3-6}
    J_0(z)=\frac1{2\pi}\int_0^{2\pi}e^{iz\cos\phi}d\phi
\end{equation}
for the Bessel function.

The integral in (\ref{3-4}) can be estimated by the method similar to
that used in Ref.~\cite{gsk}. First, reality of $\delta n$
enables one to represent Eq.~(\ref{3-4}) as a half-sum of this
expression and its complex conjugate, and to make replacements
$\bk\to-\bk$, $\eps\to-\eps$ in one of integrals. As a result we get
\begin{equation}\label{4-1}
    \delta n=\frac{V_0}{2\pi^2}\int_0^{\pi}d\eta\sin\eta\int_{-\infty}^{\infty}
    \frac{k^2J_0(kr\sin\eta\sin\chi)e^{-ikr\cos\eta\cos\chi}}
    {k^2-k_0^2-i\eps}dk,
\end{equation}
where $k$-integration takes place over the whole $k$ axis. The integrand function
has two poles
\begin{equation}\label{4-2}
    k=\pm k_0+i\eps'
\end{equation}
located in the upper complex half-plane. Hence, we can calculate this integral,
if we close the contour of integration by an infinitely large half-circle
$k=|k|e^{i\theta}$ with either $0\leq\theta\leq\pi$ or $\pi\leq\theta\leq2\pi$,
provided the contribution of these additional paths of integration vanish
as $|k|\to\infty$.

To analyze behavior of these integrals as $|k|\to\infty$, we use an asymptotic
expression for the Bessel function,
\begin{equation}\label{4-3}
    J_0(z)\approx\sqrt{\frac{2}{\pi z}}\cos\left(z-\frac{\pi}4\right),\quad z\gg1.
\end{equation}
Hence, Eq.~(\ref{4-1}) can be represented as a sum of two integrals with the
integrand functions having the exponential factors
\begin{equation}\label{4-4}
    \exp[-ikr\cos(\eta\pm\chi)-\pi/4].
\end{equation}
If $\cos(\chi\pm\eta)>0$, then this factor decays exponentially in the lower
complex $k$ half-plane, we close the contour by a lower half-circle, and since
there are no poles inside this contour, the integral vanishes in this case.
On the contrary, if
\begin{equation}\label{4-5}
    \cos(\chi\pm\eta)<0,
\end{equation}
then we close the contour of integration in the upper half-plane and both poles give non-zero
contributions into the integral. Thus, we get
\begin{equation}\label{5-1}
    \delta n=\frac{V_0}{\pi}\int_0^{\pi}k\sin\eta\cdot\sin(kr\cos\chi\cos\eta)\cdot
    J_0(kr\sin\chi\sin\eta)d\eta,
\end{equation}
where $k$ is defined by Eq.~(\ref{3-5}), i.e. we have dropped out the index for
convenience of notation.

In the far-field region $kr\gg1$ we replace the Bessel function by its
asymptotic expression (\ref{4-3}) to obtain
\begin{equation}\label{5-3}
    \delta n=\frac{V_0}{\pi}\sqrt{\frac{k\sin\eta}{\sin\chi}}\int_0^{\pi}
    \left[e^{ikr\cos(\chi+\eta)+\pi/4}+e^{ikr\cos(\chi-\eta)-\pi/4}\right]d\eta.
\end{equation}
These integrals can be calculated by the method of stationary phase.
The stationary point of the phase
\begin{equation}\label{5-4}
    s_1=k\cos(\chi+\eta)
\end{equation}
is determined by the equation $ds_1/d\eta=0$ which gives the relation
between the angles $\chi$ and $\eta$,
\begin{equation}\label{5-5}
    \tan(\chi+\eta)=-\frac{2M^2}{k^2}\sin2\eta
\end{equation}
or
\begin{equation}\label{5-6}
    \tan\chi=\frac{(1+k^2/2)\tan\chi}{M^2-(1+k^2/2)}.
\end{equation}
This expression coincides with the results obtained in the case of
2D obstacle \cite{gegk,gsk} and satisfies the condition (\ref{4-5}).
On the contrary, the second term in Eq.~(\ref{5-3}) with the phase
$s_2=k\cos(\chi-\eta)$ leads to the relation between $\chi$ and
$\eta$ which is excluded by  (\ref{4-5}). Hence we take into account
the first term only and reduce this integral into the Gaussian one
around the vicinity of the stationary point. As a result we obtain the
following distribution of the density in the wave pattern,
\begin{equation}\label{5-7}
    \delta n=\frac{2V_0}{\pi r}\frac{\{[M^2(M^2-2)\cos^2\eta+1][1+(4M^4/k^4)\sin2\eta]\}^{1/4}}
    {\{[2M^2\cos^2\eta-1][1+(4M^2/k^2)\cos2\eta+(12M^4/k^4)\sin^22\eta]\}^{1/2}}
    \cos[kr\cos(\chi+\eta)],
\end{equation}
where $\chi$ as a function of $\eta$ is determined by Eq.~(\ref{5-6}) and
$k$ is defined by Eq.~(\ref{3-5}).

The geometric form of the wave crest surfaces can be easily found in the
following way. Obviously, such a surface can be obtained by rotation of
its cross section by the $(x,y)$ plane around the $x$ axis. Then we find
from Eqs.~(\ref{3-2}), (\ref{3-5}), and (\ref{5-6}) the parametric formulae
for the coordinates of this cross section:
\begin{equation}\label{6-1}
    x=\frac{4s}{k^3}\cos\eta(1-M^2\cos2\eta),\quad
    y=\frac{4s}{k^3}\sin\eta(2M^2\cos^2\eta-1),
\end{equation}
where $s=kr\cos(\chi+\eta)$ is the phase constant along the crest line.
These formulae are identical to ones obtained in 2D case \cite{gegk,gsk}
which is natural since the Bogoliubov dispersion law for linear waves is
the same for both two and three dimensions. However, the amplitude of
waves as a function of the distance $r$ and the polar angle $\chi$
(or $\eta$) in 3D theory differs from that in the 2D case; now it decays
with $r$ as $r^{-1}$ to satisfy the energy conservation law.

As is clear from Eq.~(\ref{3-5}), the wave pattern (\ref{6-1})
corresponds to the range of the parameter $\eta$
\begin{equation}\label{6-2}
    -\arccos(1/M)\leq\eta\leq\arccos(1/M)
\end{equation}
with the coordinates located inside the Mach cone defined by the relation
\begin{equation}\label{6-3}
    \sin\chi_M=\frac1M.
\end{equation}
In particular, the small values of $\eta$ correspond to the waves located in front of
the obstacle,
\begin{equation}\label{6-4}
    x\cong-\frac{s}{2\sqrt{M^2-1}}+\frac{(2M^2-1)s}{4(M^2-1)^{3/2}}\eta^2,\quad
    y\cong\frac{(2M^2-1)s}{2(M^2-1)^{3/2}}\eta,
\end{equation}
i.e. the wave crest lines take here a parabolic form
\begin{equation}\label{6-5}
    x(y)\cong-\frac{s}{2\sqrt{M^2-1}}+\frac{(M^2-1)^{3/2}}{(2M^2-1)s}y^2.
\end{equation}
The boundary values $\eta=\pm\arccos(1/M)$ correspond to the lines
\begin{equation}\label{6-6}
    \frac{x}y=\pm\sqrt{M^2-1},
\end{equation}
i.e., far from the obstacle, they approach  the straight lines
parallel to the Mach cone (\ref{6-3}). In the region in front of the obstacle where
$y=z=0,$ $x<0$, we have $\eta=0$, hence
\begin{equation}\label{7-1}
    k=2\sqrt{M^2-1}
\end{equation}
and the wavelength
\begin{equation}\label{7-2}
    \la=\frac{2\pi}k=\frac{\pi}{\sqrt{M^2-1}}
\end{equation}
is constant. Equation (\ref{5-7}) reduces here to a simple formula
\begin{equation}\label{7-3}
    \delta n=\frac{2V_0}{\pi|x|}\sqrt{\frac{(M^2-1)(4M^2-1)}{(2M^2-1)(8M^2-1)}}
    \cos(2\sqrt{M^2-1}\,x).
\end{equation}

The formulae greatly simplify also in a highly supersonic limit and not
too close to the Mach cone when $M\cos\eta\gg1$. In this case Eq.~(\ref{5-1}) yields
\begin{equation}\label{7-4}
    \chi\cong\pi-2\eta-\frac{\sin2\eta}{2M^2}
\end{equation}
and even the leading order approximation $\chi\cong\pi-2\eta$ gives good enough
approximation in the most important region of the wave pattern. In particular,
we get the expressions for the wave crest line
\begin{equation}\label{7-5}
    x=\frac{s}{2M}(\tan^2\eta-1),\quad y=\frac{s}M\tan\eta,
\end{equation}
that is
\begin{equation}\label{7-6}
    x(y)\cong-\frac{s}{2M}+\frac{M}{2s}y^2
\end{equation}
which is the limit $M\gg1$ of Eq.~(\ref{6-5}). The density disturbance
(\ref{5-7}) takes the form
\begin{equation}\label{7-7}
    \delta n\cong\frac{V_0}{\pi r}\cos[M(r-x)],\quad M\cos\eta\gg1,
\end{equation}
and in front of the obstacle where $x=-r=-|x|$ it corresponds to the limit $M\gg1$
of Eq.~(\ref{7-3}).

\section{Numerical simulations and discussion}

In our numerical simulations the GP equation
\begin{equation}\label{8-1}
    i\hbar\frac{\prt\psi}{\prt t}=-\frac{\hbar^2}{2m}\Delta\psi+V(\br,t)\psi+
    NU_0|\psi|^2\psi,
\end{equation}
where
\begin{equation}\label{8-2}
    U_0=4\pi\hbar^2a_s/m
\end{equation}
is the effective interatomic coupling constant, $a_s$ being the $s$-wave
scattering length of atoms, $N$ is the number of atoms in the condensate,
so that $\psi$ is normalized to unity, was transformed to non-dimensional
units in the following way. We take some $a_0=\sqrt{\hbar/m\om_x}$ as a
unit of length and $\om_x^{-1}$ as a unit of time (if the BEC is confined in a
parabolic trap then $a_0$ has a meaning of the ``oscillator length'' and
$\om_x$ of the oscillator frequency along $x$ axis) and introduce
\begin{equation}\label{8-3}
    \widetilde{t}=t\om_x,\quad \widetilde{\mathbf{r}}=\mathbf{r}/a_0,
    \quad \widetilde{\psi}=\psi\cdot a_0^{3/2},\quad
    \widetilde{V}=V/(m\om_x^2a_0^2),
    \quad g=4\pi a_sN/a_0,
\end{equation}
so that the non-dimensional GP equation takes the form
\begin{equation}\label{8-5}
    i\frac{\prt\psi}{\prt t}=-\Delta\psi+V(\br,t)\psi+g|\psi|^2\psi
\end{equation}
with tildes omitted for convenience of the notation.

In the current simulations the BEC was confined in a cubic box $-10\leq x,y,z\leq 10$
and had practically uniform undisturbed distribution of density
$n_0=|\psi|^2=1.5714\cdot 10^{-4}$ except for a narrow region at the boundary of the box.
The other parameters have been chosen so that $g=8000$, the sound velocity
$c_s=\sqrt{gn_0}=1.1212$, and the healing length $\xi=1/(\sqrt{2}c_s)=0.6307$.
The potential of the obstacle was represented by a spherical ball with the radius $a_{ball}=0.125$
(which is less than the healing length) and the repulsive uniform potential equal to
$V_{ball}=100$ inside the sphere. Velocity of the ball corresponds to the Mach number
equal to $M=3$. In our simulations we have used the method of lines with spatial
discretization by Fourier pseudospectral method and time integration by adaptive Runge-Kutta
method of order 2 and 3 (RK23).

The resulting wave patterns is shown in Fig.~2.
We have found that it is axially symmetric, as it was supposed,
and the $(x,y)$ cross sections
of the wave crest surfaces agree very well with the analytical curves
shown by dashed lines and corresponding
to Eqs.~(\ref{6-1}). Oscillations of the density in front of the obstacle
as a function of the $x$ coordinate is shown in Fig.~3 and it is compared with
the analytical expression (\ref{7-3}). Again good agreement is observed.
Thus, the wave pattern located outside the Mach number is described quite
satisfactory by the developed here theory.
\begin{figure}[bt]
\begin{center}
\includegraphics[width=10cm,clip]{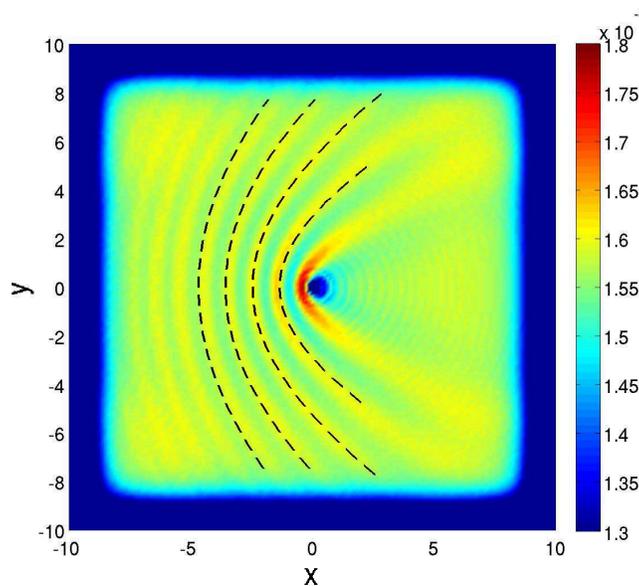}
\caption{(color online) Numerically simulated wave pattern generated by a spherical obstacle
moving through a Bose-Einstein condensate. The parameters of the BEC are indicated
in the text. The analytical wave crest lines are shown by dashed lines.
They are plotted according to Eqs.~(\ref{6-1}).
}
\end{center}
\end{figure}

\begin{figure}[bt]
\begin{center}
\includegraphics[width=10cm,clip]{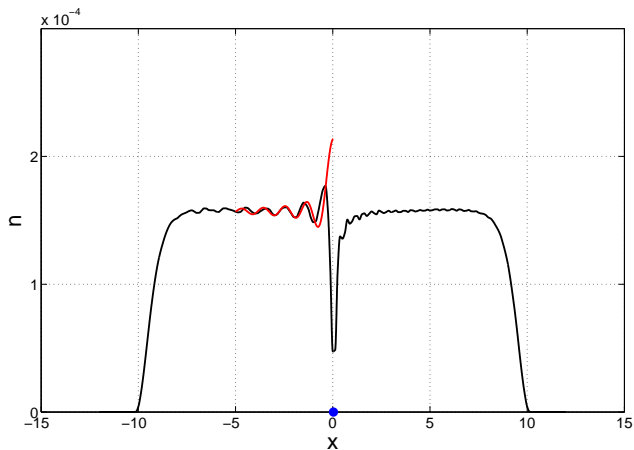}
\caption{(color online) Oscillatory structure in front of the obstacle obtained
by numerical simulations
(solid line) and analytically (Eq.~(\ref{7-3}); red line). Position of the obstacle
is shown by a blue dot.}
\end{center}
\end{figure}

Since the size of the obstacle is much less than the healing length, there were
no formation of vortex rings located inside the Mach cone. Just these
structures attracted earlier much
attention in the study of the loss of superfluidity in a subsonic motion of
obstacles (see, e.g., \cite{br-2000} and references therein) when stationary
``ship waves'' patterns do not exist---in subsonic case only time-dependent linear
waves can be generated due to the switching on the obstacle potential \cite{bks-08}
or a change of the obstacle velocity. 
Similar vortex-antivortex pairs are also
generated in the 2D case where they align along
straight lines as the velocity of the obstacle grows and above some
critical value of velocity one can see the formation of
oblique dark solitons attached at one their end to the obstacle and decaying into
vortices at the other end. One may suppose that if the size of the obstacle
exceeds the healing length, then in the 3D case formation of
``conical solitons'' would take place above some critical velocity. However,
the study of this problem is outside the scope of the present paper.

In conclusion, we have studied the formation of linear wave pattern generated
by a 3D small obstacle moving with a supersonic velocity through a uniform
condensate. Analytical formulae for the wave crest lines and dependence of the
amplitude of the density oscillations on the distance from the obstacle
are confirmed by numerical simulations. This theory essentially extends
previous calculations of the ``drag force'' and provides a more detailed
picture of the process of Cherenkov radiation of Bogoliubov excitations
in rarefied Bose condensates.

\subsection*{Acknowledgments}

This work was supported partially by National Center for Theoretical Sciences
(NCTS), Hsinchu, Taiwan, and Taida Institute for Mathematical Sciences
(TIMS), Taipei, Taiwan.
A.P.I. was supported by the Academy of Finland (Project No. 213362), 
and partially by RFBR 06-01-00117. A.M.K. thanks RFBR
for partial support.


\begin{thebibliography}{99}

\bibitem{landau1} L.D. Landau, J. Phys. USSR, {\bf 5,} 71 (1940).

\bibitem{landau2} L.D. Landau, J. Phys. USSR, {\bf 11,} 91 (1947).

\bibitem{feynman} R.P. Feynman,  in {\it Progress in Low Temperature
Physics} Vol. I, ed C.J. Gorter, p. 17 (Amsterdam: North-Holland) (1955).

\bibitem{km-01} D.L. Kovrizhin and L.A. Maksimov, Phys. Lett.
A {\bf 282,} 421 (2001).

\bibitem{ap-04} G.E. Astrakharchik and L.P. Pitaevskii, Phys. Rev. A {\bf 70,}
013608 (2004).

\bibitem{cornell05} E.A. Cornell, ``Conference on
Nonlinear Waves, Integrable Systems and their Applications",
(Colorado Springs, June 2005);
http://jilawww.colorado.edu/bec/papers.html.

\bibitem{carusotto} I. Carusotto, S.X. Hu, L.A. Collins, and A. Smerzi,
Phys. Rev. Lett. {\bf 97,} 260403 (2006).

\bibitem{wmca-99} T. Winiecki, McCann, and C.S. Adams, Phys. Rev. Lett.
{\bf 82,} 5186 (1999).

\bibitem{gegk}  Yu.G. Gladush, G.A. El, A. Gammal, A.M. Kamchatnov, Phys. Rev. A
{\bf 75,} 033619, (2007).

\bibitem{gsk}  Yu.G. Gladush, L.A. Smirnov, and A.M. Kamchatnov, J. Phys. B:
Mol. Opt. Phys. {\bf 41,} 165301 (2008).

\bibitem{egk1} G.A. El, A. Gammal, and A.M. Kamchatnov, Phys. Rev. Lett.
{\bf 97,} 180405 (2006).

\bibitem{egk2} G.A. El, Yu.G. Gladush, and A.M. Kamchatnov,
J. Phys. A: Math. Theor. {\bf 40,} 611 (2007).

\bibitem{kp08} A.M. Kamchatnov and L.P. Pitaevskii, Phys. Rev. Lett.
{\bf 100,} 160402 (2008).

\bibitem{two-comp} Yu.G. Gladush, A.M. Kamchatnov, Z. Shi, P.G. Kevrekidis,
D.J. Frantzeskakis, B.A. Malomed, arXiv:0811.1891.

\bibitem{sh-08} R. G. Scott and D. A. W. Hutchinson, Phys. Rev. A
{\bf 78,} 063614, (2008).

\bibitem{br-2000} N.G. Berloff and P.H. Roberts, J. Phys. A {\bf 33}, 4025 (2000).

\bibitem{bks-08} B.B. Baizakov, A.M. Kamchatnov, and M. Salerno,
J. Phys. B {\bf 41,} 215302 (2008).

\end{thebibliography}
\end{document}